\documentclass[10pt,twocolumn,showpacs,english,nofootinbib,longbibliography]{revtex4-1}
\usepackage{amssymb,amsmath}
\usepackage{amsbsy}
\usepackage{float}
\usepackage{times}
\usepackage[T1]{fontenc}
\usepackage[latin9]{inputenc}
\usepackage{amsbsy}
\usepackage{graphicx}
\usepackage{esint}
\usepackage{array}
\usepackage[unicode,breaklinks]{hyperref}
\hypersetup{
    unicode=true,
    a4paper=true,
    plainpages=false, 
    colorlinks=true,
    linkcolor=blue,
    citecolor=blue,
    filecolor=black,
    urlcolor=blue
}
\usepackage{verbatim}
\newcommand{\bk}{\mathbf{k}}

\newcommand{\bx}{\mathbf{x}}

\newcommand{\F}{\mathcal{F}}
\newcommand{\M}{\mathcal{M}}
\newcommand{\N}{\mathcal{N}}
\newcommand{\V}{\mathcal{V}}
\newcommand{\Z}{\mathcal{Z}}

\newcommand{\nh}{\hat{n}}
\newcommand{\Ah}{\hat{A}}
\newcommand{\Bh}{\hat{B}}
\newcommand{\Fh}{\hat{F}}

\newcommand{\Hh}{\hat{H}}
\newcommand{\Nh}{\hat{N}}

\newcommand{\Qh}{\hat{Q}}
\newcommand{\Vh}{\hat{V}}
\newcommand{\Zh}{\hat{Z}}
\newcommand{\rhoh}{\hat{\rho}}
\newcommand{\psih}{\hat{\psi}}
\newcommand{\psihd}{\psih^\dagger}

\newcommand{\Ft}{\tilde{F}}

\newcommand{\Hint}{\Hh_{\mathrm{int}}}
\newcommand{\HBragg}{\Hh_\mathrm{Bragg}}
\newcommand{\av}[1]{\langle #1 \rangle}
\newcommand{\alh}{\hat{\alpha}}
\newcommand{\hc}{\mathrm{h.c.}}
\newcommand{\cc}{\mathrm{c.c.}}
\newcommand{\intinf}{\int_{-{\infty}}^{\infty}}
\newcommand{\tdiff}[2]{\dfrac{d #1}{d #2}}
\newcommand{\z}{\mathcal{Z}}

\usepackage{babel}

\begin{document}

\title{Spin-dependent Bragg spectroscopy of a spinor Bose gas} 
 
\author{D.~Baillie} 
\affiliation{Dodd-Walls Centre for Photonic and Quantum Technologies, Department of Physics, University of Otago, Dunedin, New Zealand} 
\author{P.~B.~Blakie}  
\affiliation{Dodd-Walls Centre for Photonic and Quantum Technologies, Department of Physics, University of Otago, Dunedin, New Zealand}

\begin{abstract}
We develop a general theory of spin-dependent Bragg spectroscopy for spinor Bose-Einstein condensates. This spectroscopy involves using a density and spin-coupled optical probe to excite the system. We show that within the linear response regime the momentum or energy transferred by the probe is determined by a set of density and spin-density dynamic structure factors. We derive a set of $f$-sum rules that provide rigorous constraints for the first energy moments of these structure factors. As an application we compute the dynamic structure factors for cases within all four distinct phases of a spin-1 condensate using Bogoliubov theory. Our results demonstrate that spin-dependent Bragg spectroscopy can be used to selectively investigate the various phonon and magnon excitation branches and will be a useful tool for advancing our understanding of spinor condensates.
\end{abstract}


\maketitle

\section{Introduction}
A spinor Bose-Einstein condensate is a quantum degenerate gas in which the atoms are able to access their spin degrees of freedom. In addition to being a superfluid, this system can exhibit various types of magnetic ordering depending on the interaction parameters and externally applied fields (e.g.~see \cite{Kawaguchi2012R}). There has been significant progress in the field with the development of novel methods for measuring aspects of spin order in the condensate (e.g.~see \cite{Higbie2005a,Liu2009a,Liu2009b,Zhang2009a}).
 It is of interest to understand the rich excitation spectrum of this system, which should exhibit   phonon and various magnon branches. These excitations reveal aspects of the ordered phases (e.g.~broken symmetries \cite{Murata2007a,Symes2014a}), and can play an important role in the dynamics that occurs when the system transitions between magnetic phases (e.g.~see \cite{Saito2007a}). 
 Also, it has recently been shown that the excitation spectrum of a spinor condensate exhibits the elusive quantum mass acquisition process (where a massless quasiparticle becomes massive by quantum corrections) \cite{Phuc2014a}. There are theoretical proposals for spinor condensate spectroscopy \cite{Tokuno2013a} and fluctuation measurements \cite{Symes2014b}  that are sensitive to the nature of the excitations, but are not energy and momentum resolved. One important recent step made in experiments has seen the measurement of the long wavelength magnon dispersion relation in a particular phase of a spin-1 condensate using an interferometric technique \cite{Marti2014a}. In this paper we consider an alternative scheme for probing a spinor condensate with a Bragg spectroscopy technique that is energy,  momentum and spin sensitive. 

Bragg spectroscopy is a commonly used experimental tool for probing a wide range of properties in (non-spinor) condensates, such as dynamic and static structure  factors \cite{Stamper-Kurn1999a}, the momentum distribution and coherence \cite{Stenger1999b,Hugbart2007a},  and to detect vortices \cite{Blakie2001a,Muniz2006a} and roton-like features \cite{Blakie2012a,Ha2015a} (also see \cite{OzeriR2005a,Thywissen2011a}).  The basic idea in Bragg spectroscopy is to use a stimulated two-photon process to scatter atoms in an energy and momentum resolved manner, with the system response subsequently determined by measuring the amount of excitation (e.g.~momentum or energy transfer to the system).  Bragg spectroscopy is in some sense analogous to neutron scattering, which is also used to measure the dynamic structure factor in condensed matter systems and was used to confirm the energy-momentum spectrum of excitations of superfluid helium \cite{Yarnell1959a}. Because neutrons have a magnetic moment they are also able to probe magnetic degrees of freedom, and neutron scattering was used to make the first direct measurement of the magnon dispersion relation in magnetite \cite{Brockhouse1957a}. Similarly, by performing Bragg spectroscopy using a two-photon process that is dependent on the spin state of the atoms, it will be possible to probe the magnetic structure of the condensate and its excitations.
 The polarization of light can be used to effect such a spin-dependent coupling \cite{Cohen-Tannoudji1972a,Deutsch1998a}, and was basis of a proposal to measure the structure factor of two-component gases. Finite detuning of the two-photon process can also be used to arrange a spin-dependence in Bragg spectroscopy, and has been successfully applied in 
recent experiments with a strongly interacting (two-component) Fermi gas \cite{Hoinka2012a}. While there has been prior work in aspects of the spin-dependent spectroscopy (or related dynamic structure factors) in  Fermi gases (e.g.~see \cite{Bruun2006a,Combescot2006a,Guo2010a,Hu2012a,Astrakharchik2014a}), the case of a spinor condensates has gone largely unexplored.

The spin dependent Bragg spectroscopy technique  we propose involves two steps to make a measurement: 
\textbf{(i)} The system is excited by a weak spin-dependent Bragg probe of adjustable wavevector and frequency, i.e.~moving optical dipole potential that has a vectorial coupling to the atomic hyperfine sub-levels.
\textbf{(ii)} The momentum or energy imparted to the system is subsequently measured to quantify the system response. Such measurements are then repeated on an identically prepared spinor condensate for a range of Bragg frequencies, and often different wavevectors. We note that this technique only differs from the usual Bragg spectroscopy technique for scalar condensates by the spin-dependence of the excitation probe. As noted above, there are several ways to produce such a spin-dependent coupling. In this work our focus is upon developing a theoretical description for this scheme in terms of linear response theory, and to relate the measured properties to density and spin-density dynamic structure factors. An important result is that these dynamic structure factors are often dominated by different excitation branches in the spinor condensate. Thus with an appropriately chosen spin-dependent Bragg probe it will be possible to selectively probe the density (or phonon) branch, or one of the magnon branches.

The outline of our paper is as follows. We begin by defining the Bragg probe and observables we consider. We develop the formalism in a general manner for bosonic atoms of arbitrary integer spin $f$ interacting with short range rotationally invariant interactions. We use linear response theory to relate the response of the system observable after weak excitation to equilibrium dynamical correlation functions of the system. We choose to characterize these correlation functions in terms of a set of density and spin-density dynamic structure factors. We then focus on understanding their properties. Using commutations relations of the full many-body Hamiltonian, we derive rigorous $f$-sum rules that specify the first frequency moment of the dynamic structure factors as a function of the wavevector of excitation. We formulate the dynamic structure factors using meanfield (Bogoliubov) treatment of the excitations suitable for practical calculations. Finally we present the results of calculations to illustrate the behavior of the dynamic structure factors for a spin-1 condensate in the four distinct magnetic phases accessible to this system. 
 
\section{Formalism}
\subsection{Hamiltonian}
Here our interest is in spin-$f$ bosons described by the $(2f+1)$-component bosonic field operator  $\psih_m(\bx)$ with $m\in\{-f,-f+1,\ldots,f\}$. It is useful to define the operators
\begin{align}
\hat{n}(\bx)&= \sum_{m}\psih^\dagger_m(\bx)\psih_m(\bx),\\
\Fh_\mu(\bx)&= \sum_{mm'}(f_\mu)_{mm'}\psih^\dagger_m(\bx)\psih_{m'}(\bx),\\
\hat{\N}_{\mu\mu'}(\bx)&= \sum_{mm'}\frac{1}{2}(f_\mu f_{\mu'}+f_{\mu'} f_\mu)_{mm'}\psih^\dagger_m(\bx)\psih_{m'}(\bx),\label{Eq:nematicden}
\end{align}
which describe the total, spin and nematic density, respectively, with $\mu\in\{n,x,y,z\}$ where $f_\mu$ is the spin-$f$ matrix augmented by $(f_n)_{mm'}=\delta_{mm'}$ so that $\Fh_n(\bx)=\hat{\N}_{nn}(\bx)=\nh(\bx)$ and $\hat{\N}_{n\mu}(\bx)=\Fh_\mu(\bx)$. 
The Hamiltonian for a dilute spin-$f$ Bose gas the can be written in the form \cite{Kawaguchi2012R} (also see \cite{Ho1998a,Ohmi1998a})
\begin{align}
    \Hh &= \Hh_0+\Hh_Z +  \Hint,\label{Hspinor}
    \end{align}
    where
    \begin{equation}
    \Hh_0 =  \int d\bx\sum_{m} \psih_m^\dagger\left[- \frac{\hbar^2 \nabla^2}{2M} +\frac{M}{2}\sum_{i \in \{x,y,z\}}\omega_i^2x_i^2\right]\psih_m,
    \end{equation}
    is the spin-independent single particle Hamiltonian with harmonic confinement, with $M$ the atomic mass and $\omega_i$ the trap frequency in direction $i$. A uniform  magnetic field is applied along $z$ causing a Zeeman term
    \begin{equation} 
    \hat{H}_Z= \int d\bx\, \left[-p\Fh_z(\bx)+q\hat{\N}_{zz}(\bx)\right],
    \end{equation}
where  $p$ and $q$ are the linear and quadratic Zeeman energies, respectively. Finally, the spinor interaction Hamiltonian can be written in the form
    \begin{align}  
        \Hint &= \sum_{\F=0,2,\dots,2f} \Hint^{(\F)},\label{spinorints}\\
     \Hint^{(\F)}     &=  \frac{g_\F}{2} \int d\bx\sum_{\M=-\F}^\F \Ah^\dag_{\F\M}(\bx) \Ah_{\F\M}(\bx),\\
     \Ah_{\F\M}(\bx) &= \sum_{mm'=-f}^f  \av{\F\M|mm'} \psih_m(\bx)\psih_{m'}(\bx),
\end{align}
where $\Hint^{(\F)}$ is the interaction Hamiltonian for two bosons with total spin $\F$, $g_\F=4\pi\hbar^2a_\F/M$ is the coupling constant for the spin-$\F$ channel with $s$-wave scattering length $a_\F$, $\Ah_{\F\M}(\bx)$ is the irreducible operator annihilating a pair of bosons at $\bx$ and $\av{\F\M|mm'}\equiv\av{\F,\M|f,m;f,m'}$ is the Clebsch-Gordan coefficient \cite{Kawaguchi2012R}.
\subsection{Spin- and density-dependent Bragg probe}
We consider a pair of laser beams in a stimulated Bragg scattering configuration that are well detuned from atomic resonance, so that spontaneous emission can be neglected. This gives rise to a moving Bragg potential experienced by the atoms with a wave vector  $\bk$ set by the difference of the wave vectors between the two beams, moving at a speed $\omega/k$, where $\omega$ is the difference in frequencies of the two beams. In what follows we will take the Bragg potential to be oriented along $z$, i.e.~$\mathbf{k}=k\hat{z}$. Using the polarisation of the light \cite{Cohen-Tannoudji1972a,Deutsch1998a} or adjusting the detuning of the beams from the excited states it is possible to realize a spin-dependent coupling to the atomic field (see  \cite{Carusotto2006a,Hoinka2012a})  
\begin{align}
    \HBragg &= \frac{V}{2}\Bh^\dagger_\bk e^{-i\omega t+\eta t} + \hc,  \label{eq:HBragg}
\end{align}
where the strength of the Bragg potential, $V$, is dependent on the laser beam intensity and detuning, and $\eta\to0^+$.
We have introduced the spin-dependent density fluctuation operator
\begin{align}
\Bh_{\bk}\equiv\int d\bx\,e^{-i\bk\cdot\bx}\Bh(\bx),
\end{align}
which characterizes  the coupling of the probe to the atoms at the Bragg wave vector $\mathbf{k}$, with 
\begin{align}
\Bh(\bx) = \sum_{mm'}B_{mm'}\psih^\dagger_m(\bx)\psih_{m'}(\bx),
\end{align}
being a general spin-density of the system where $B_{mm'}$ is a dimensionless $3\times3$ coupling matrix. 
For the case of standard (spin-independent) Bragg spectroscopy, the probe only couples to the total density, and $B_{mm'} = \delta_{mm'}$.  Most methods considered for producing a spin-dependent coupling do not result in a potential that  couples to only a single component of spin density, but have both density and spin-density terms, of a form dependent on the scheme employed. Here we will make a reasonably general choice for the coupling matrix
\begin{align}
    B_{mm'} &= \sum_{\mu\in\{n,x,y,z\}} B_\mu (f_\mu)_{mm'}, \label{e:Bdef} 
\end{align}
so that it couples to the density and spin-density.   
This choice encompasses two important cases. The polarisation dependent proposal for spin-$\frac12$ Fermi gases in \cite{Carusotto2006a} is obtained by taking $\mathbf{B}=(B_n,0,0,i B_z)$, and the detuning based scheme employed in \cite{Hoinka2012a} for a spin-balanced Fermi gas is obtained by taking $\mathbf{B}=(B_n,0,0,B_z)$ for real $B_n$, $B_z$.  Other cases will be possible, e.g.~by varying the polarization of the light fields (e.g.~see \cite{Deutsch1998a}).

\subsection{Response observable}
 Complementing the Bragg probe is the scheme used to observe the effect on the condensate. Here we present two observables suitable for this purpose. We emphasise that both of these observables are commonly used for measuring the response of spinless Bose gases  (e.g.~see \cite{Stenger1999b,Stamper-Kurn1999a,Brunello2001a,Blakie2002a,Steinhauer2002a,Stoferle2004a,Clement2009a}) and two-component Fermi gases (e.g.~see \cite{Veeravall2008a,Kuhnle2011a,Hoinka2012a}), and will be applicable to the spinor gas.
  
 \subsubsection{Imparted momentum}\label{SecPimpart}
 The first observable we examine is the rate of momentum imparted to the condensate, which is initially at rest.  For our case with the Bragg wave vector  along $z$ we consider the $z$-component of momentum 
\begin{align}
    \hat{P}_z &= \frac{\hbar}{i}\int d\bx\,\sum_{m}\psih^{\dagger}_m(\bx)\frac{\partial}{\partial z}\psih_m(\bx),
\end{align}
from which the observable is obtained by evaluating the Heisenberg equation of motion for $\hat{P}_z$ under the influence of the perturbation 
\begin{align}
\frac{d P_z(t)}{dt} &= \frac{1}{i\hbar}\langle [\hat{P}_z,\Hh+\HBragg]\rangle,\\
&=-M\omega_z^2Z -\frac{ik V }2 (\av{\delta \Bh^\dagger_\bk} e^{-i\omega t}-\cc ),\label{Eq:Pzrate}
\end{align}
where 
\begin{equation}\delta \Bh_{\bk}\equiv\int d\bx\,e^{-i\bk\cdot\bx}\left[\Bh(\bx)-\langle \Bh(\bx)\rangle_0\right],\label{e:deltaB}\end{equation}%
$\av{\Bh(\bx)}_0$ signifies equilibrium expectation, and $Z\equiv  \int d\bx\,z\langle\hat{n}(\bx)\rangle$ is the $z$-component of the centre of mass coordinate.
 In experiments $Z$ is initially zero, and can be assumed to remain small if the Bragg pulse is sufficiently short compared to the trap period in the scattering direction. The value of  $P_z$ (and hence the rate $\dot{P}_z$) is then typically determined by allowing the system to freely expand (i.e.~turning the harmonic confinement off) at the conclusion of the Bragg pulse and measuring the centre of mass displacement of the system.
 
 \subsubsection{Imparted energy}\label{SecEimpart}
An alternative observable is the rate at which energy is imparted to the cloud by the perturbation (e.g.~see \cite{Stoferle2004a,Kuhnle2011a}). In this case the observable is the unperturbed Hamiltonian operator for the system
\begin{align}
\frac{d E(t)}{dt} &= \frac{1}{i\hbar}\langle [\Hh,\Hh+\HBragg]\rangle , \\
&= -\frac{V}{2}\frac{  d\av{\delta \Bh^\dagger_{\bk}}}{dt} e^{-i\omega t} + \cc,\label{Eq:Erate}
\end{align}
which can be measured by holding the system in trap until it rethermalizes, and then measuring the increase in temperature.  

\subsection{Linear response treatment of fluctuations\label{s:linresponse}}  
We assume that the system starts in an equilibrium state of the (unperturbed) Hamiltonian, $\hat{H}$, and using  linear response theory we obtain results for the evolution of the fluctuations $\langle \delta \Bh_{\bk}\rangle$ 
(see Appendix \ref{s:linresp}).
This analysis shows that the rate of change of the $z$ component of momentum and the energy are
\begin{align}
    \frac{d P_{z}(t)}{d t} &=  -M\omega_z^2Z  + \frac{k V^2}{2} \! \int d\omega' \frac{\sin[(\omega\!-\!\omega')t]}{\omega - \omega'}\notag\\
    & \times\left[S_{B,B^\dagger}(\bk,\omega')-S_{B^\dagger,B}(\bk,-\omega')\right],\label{EqPresponse}\\
    \frac{d E(t)}{dt}   &= \frac{V^2}{2}\int d\omega'\,\omega'\frac{\sin[(\omega-\omega')t]}{\omega-\omega'}\notag\\
    &\times[S_{B,B^\dagger}(\bk,\omega')-S_{B^\dagger,B}(\bk,-\omega')].\label{EqEresponse} 
\end{align}
Here we have introduced the generalized dynamic structure factor  
 \begin{equation}
     S_{Q,R}(\bk,\omega)=\sum_{mn}\frac{e^{-\beta E_m}}{\hbar\Z}\av{m|\delta\Qh_{\bk}|n}\av{n| \delta\hat{R}_\bk|m}\delta(\omega-\omega_{nm}),\label{Eq:Smu}
  \end{equation}
where $Q$ and $R$ represent spatially dependent operators $\Qh(\bx)$, $\hat{R}(\bx)$, with $\delta \Qh_{\bk}$ defined as in \eqref{e:deltaB}, $E_m$ is the energy of the unperturbed Hamiltonian with respective eigenstate $|m\rangle$,  $\omega_{mn}=(E_m-E_n)/\hbar$, $\beta=1/k_BT$ the inverse temperature,  and $\Z=\sum_me^{-\beta E_m}$.  

The Bragg excitation operator $\hat{B}$ has density and spin density parts [see Eq.~(\ref{e:Bdef})], and it is convenient to express the dynamic structure factor $S_{B,B^\dag}$ as
  \begin{align}
      S_{B,B^\dag}(\bk,\omega) &=  \sum_{\mu\mu'} B_\mu B^*_{\mu'}\, S_{\mu\mu'}(\bk,\omega), \label{e:SBterms}
\end{align}
where we have introduced the elementary dynamic structure factor
\begin{align}
S_{\mu\mu'}(\bk,\omega)&\equiv S_{F_\mu,F^\dag_{\mu'}}(\bk,\omega).\label{Eq:SFmumud}
\end{align}
$S_{nn}$ describes the ``density channel'' (following the terminology of \cite{Hoinka2012a}). For $\mu,\mu'\in\{x,y,z\}$,  $S_{\mu\mu'}$ describes the $\mu\mu'$-component of the ``spin channel'',  and $S_{n\mu}$ the spin-density channel. By adjusting the Bragg probe (e.g.~varying the detuning, as in \cite{Hoinka2012a}) it may be possible to probe the spin or density channels separately, or to infer them from the response measurements made with two different Bragg operators $\hat{B}$. For the remainder of this paper we will focus on understanding the properties of $S_{\mu\mu'}$, since the results from any particular measurement scheme can be expressed in terms of these.
 
Alternatively, we could work in terms of dynamic susceptibility functions. The relationship between the spin-density dynamic structure factors and susceptibility tensor is briefly discussed in Appendix \ref{s:linresp}.

\subsection{Sum rules}\label{Sec:SumRules}
Useful information about the dynamic structure factors can be obtained by using the method of sum rules \cite{BECbook}, which provides an algebraic method to evaluate certain moments of the structure factor without needing to solve for the exact eigenstates.

Consider the $p^{\mathrm{th}}$ order energy moment  of the structure factor
\begin{align}
    m^p_{\mu\mu'}(\bk) &\equiv \hbar\int d\omega\,(\hbar \omega)^p S_{\mu\mu'}(\bk,\omega).
\end{align}
\subsubsection{Zero-moment: static structure factor} 
The  $p=0$ moment  defines the static structure factor, $S_{\mu\mu'}(\bk)$, conventionally defined as  $ S_{\mu\mu'}(\bk)=N^{-1}m^0_{\mu\mu'}(\bk)$, where $N$ is the number of particles in the system. The static structure factors characterise fluctuations in the system, i.e.~
\begin{align}
    S_{\mu\mu'}(\bk) &= \frac1N\av{\delta \Fh_{\mu,\bk} \delta\Fh_{\mu',-\bk}}_0. 
\end{align}
The static structure factors relate to the Fourier transform generalized (density and spin-density) pair correlation functions and have been considered for spinor condensates in Ref.~\cite{Symes2014a}. Notably, there it is shown that in the high-$k$ regime the static structure factor approaches the uncorrelated value 
\begin{equation}
     S_{\mu\mu'}(k\!\to\!\infty)=\frac{1}{N}\int d\bx \sum_{mm'}   (f_{\mu}f_{\mu'})_{mm'} \av{\psihd_m(\bx) \psih_{m'}(\bx)}_0,
\end{equation}
i.e.~for $\mu,\mu'\in\{x,y,z\}$ it is dependent on the nematic order of the system [cf.~Eq.~(\ref{Eq:nematicden})]. 
In contrast, the density structure factor has the well-known high $k$ limit $S_{nn}(k\to\infty)=1$, independent of the system state.

\subsubsection{First-moment: the $f$-sum rule} 
The  $p=1$ moment,
$m^1_{\mu\mu'}(\bk)  = \hbar^2\int d\omega\,\omega S_{\mu\mu'}(\bk,\omega)$, can be used to obtain the $f$-sum rule for this system, through a double commutator of the fluctuation operators with the Hamiltonian (as evaluated in Appendix~\ref{s:sumrules})
\begin{align}
    \Re\{m^1_{\mu\mu'}(\bk) \} &= \frac12\av{[\Fh_{\mu,\bk}, [\Hh,\Fh_{\mu',-\bk}]]}_0, \\
 &= \epsilon_\bk  \N_{\mu\mu'} + m_{Z,\mu\mu'},\label{e:sumrules} 
\end{align} 
where the only non-zero elements of $m_{Z,\mu\mu'}$ are for $\mu,\mu'\in\{x,y\}$ with 
\begin{align}
    m_{Z,\mu\mu'} &= -q\N_{\mu\mu'} + \delta_{\mu\mu'}\left\{\frac{p}2 F_z\!-\!q[2\N_{zz}\!-\!Nf(f\!+\!1)]\right\}, \label{e:m1zeeman} 
\end{align}
where we have introduced $\epsilon_\bk\equiv\hbar^2k^2/2M$, $\N_{\mu\mu'}\equiv \int d\bx\,\av{\hat{\N}_{\mu\mu'}(\bx)}_0$, and $F_\mu\equiv \int d\bx\,\av{\Fh_{\mu}(\bx)}_0$, and note that $\N_{nn}=F_n=N$, $\N_{n\mu}=F_\mu$. 
\section{Bogoliubov result for the dynamic structure factors of  a uniform spinor condensate \label{s:bog}}
In order to make quantitative predictions for the dynamic structure factors we employ a meanfield Bogoliubov treatment of the condensate and its excitations. 
We briefly review the standard theory for the uniform system (see \cite{Kawaguchi2012R} for more details) to introduce our notation for the condensate and quasiparticles. We then evaluate the density and spin-density dynamic structure factors in terms of these quantities.

\subsection{Bogoliubov theory of a uniform spinor condensate}
We consider to a uniform system in a volume $\V$ with periodic boundary conditions. We expand the field operator in a basis of plane wave modes
 \begin{equation}
 \psih_m(\bx)=\sum_{\mathbf{k}}\frac{e^{i\mathbf{k}\cdot\bx}}{\sqrt{\V}}\,\hat{a}_{m\mathbf{k}},
 \end{equation}
 with $\hat{a}_{m\mathbf{k}}$ being the operator for annihilating a particle in momentum state $\hbar\mathbf{k}$ with spin projection $m$.
Making the  Bogoliubov approximation, the field operator is separated into a zero-momentum condensate part and a non-condensate operator as
\begin{equation}
 \psih_m(\bx)=\sqrt{n}\xi_m+\hat{\delta}_m(\bx),
\end{equation}
where $n$ is the condensate total density, $\xi_m=\langle\psih_m\rangle/\sqrt{n}$ is a uniform normalised spinor (i.e.~$\sum_m|\xi_m|^2=1$) and the non-condensate operator has the property $\langle \hat{\delta}_m\rangle=0$. 

The condensate spinor is determined by minimising an energy functional (see \cite{Kawaguchi2012R}) and exhibits a rich phase diagram of ground states with different spin ordering depending on the value of the spin-dependent interaction, density and Zeeman energies. The condensate  spin and nematic moments are given by
\begin{align}
F_\mu&= N\sum_{mm'}(f_\mu)_{mm'}\xi_m^*\xi_{m'},\label{e:CondF}\\
\mathcal{N}_{\mu\mu'}&= \frac{N}{2}\sum_{mm'}(f_\mu f_{\mu'}+f_{\mu'} f_\mu)_{mm'}\xi_m^*\xi_{m'},\label{e:CondN}
\end{align}
where  $N=n\V$ and for temperatures well-below the critical temperature where the depletion is negligible, these can be taken as the system moments [e.g.~as required in the sum rules \eqref{e:sumrules}].

The non-condensate operator is approximated in a quasiparticle expansion of the form
\begin{equation}
\hat{\delta}_m(\bx)=\sum_{\mathbf{k}\ne\mathbf{0},\nu}(u^{\mathbf{k}}_{m\nu}\alh_{\mathbf{k}\nu}-{v^{-\mathbf{k}}_{m\nu}}^*\alh_{-\mathbf{k}\nu}^\dagger) 
\frac{ e^{i\bk\cdot\bx}}{\sqrt{\V}},
\end{equation}
where $\alh_{\mathbf{k}\nu}$ are quasiparticle operators with $\nu=0,1,2$ labelling the quasiparticle branch. The quantities $\hbar\omega_{\bk\nu}$ and $\{u^{\mathbf{k}}_{m\nu},v^{\mathbf{k}}_{m\nu}\}$ are the quasiparticle energies and amplitudes, respectively, and can be obtained by diagonalizing  
a $[2(2f+1)]\times[2(2f+1)]$ matrix  for each value of $\bk$ (e.g.~see Secs.~5.1 and 5.2 of Ref.~\cite{Kawaguchi2012R}).  

\subsection{Dynamic structure factors} 

We  evaluate the dynamic structure factors in terms of the condensate and quasiparticles as
\begin{align} 
  S_{\mu\mu'}(\bk,\omega)  =  \frac{N}{\hbar}&\sum_{\nu} \bigl[\delta\Ft_{\mu,\bk\nu}\delta\Ft^*_{\mu',\bk\nu}(\bar{n}_{\mathbf{k}\nu}\!+\!1)\delta(\omega-\omega_{\bk\nu})\notag\\
  &  +\delta\Ft^*_{\mu,\bk\nu}\delta\Ft_{\mu',\bk\nu}\bar{n}_{\mathbf{k}\nu}\delta(\omega+\omega_{\bk\nu})     \bigr], \label{e:Bogsmumud}
   \end{align}
   where $\bar{n}_{\mathbf{k}\nu}=[\exp(\beta\hbar\omega_{\mathbf{k}\nu})-1]^{-1}$ is the thermal occupation of the quasiparticle mode, and we have defined
 \begin{align}
     \delta\tilde{F}_{\mu,\bk\nu} &= \sum_{mm'}   (f_\mu)_{mm'}\left(\xi^*_m u^\bk_{m' \nu} - v^\bk_{m \nu} \xi_{m'}\right). \label{e:tildeFdef} 
\end{align}
We recall that the measurement observables  [Eqs.~\eqref{EqPresponse} and \eqref{EqEresponse}] relate to the general dynamic structure factor $S_{B^\dagger,B}$, which can be immediately determined from the above results 
using \eqref{e:SBterms}. It is worth noting that the thermal factors that appear in the dynamic structure factor (i.e.~$\bar{n}_{\mathbf{k}\nu}$) cancel because the observable depends on the combination $S(\bk,\omega)-S(\bk,-\omega)$  [e.g.~see Eqs.~(\ref{EqPresponse}) and (\ref{EqEresponse})]. Thus the system response to Bragg spectroscopy is independent of temperature.

\section{Results for a spin-1 condensate}
 
We now illustrate the behaviour of the structure factors based on the formalism of Sec.~\ref{s:bog} for a spin-1 condensate where the Hamiltonian reduces to
 \begin{align}
 \hat{H} =&    \int d\bx\,\left[\sum_m\psih_m^\dagger\left(-\frac{\hbar^2\nabla^2}{2M}\right)\psih_m -p\Fh_z+q\hat{\N}_{zz} \right] \nonumber  \\
 &+\int d\bx\,:\frac{c_0}{2}\hat{n}^2+\frac{c_1}{2}\sum_{i\in\{x,y,z\}}\hspace{-3mm}\Fh_i^2:,\label{Hspin1}
 \end{align}
 where $::$ indicates normal ordering, and $c_0=\frac{1}{3}(g_0+2g_2)$, $c_1=\frac{1}{3}(g_2-g_0)$ are the density and spin-dependent interactions, respectively.

 \begin{figure}
\includegraphics[width=3.5in]{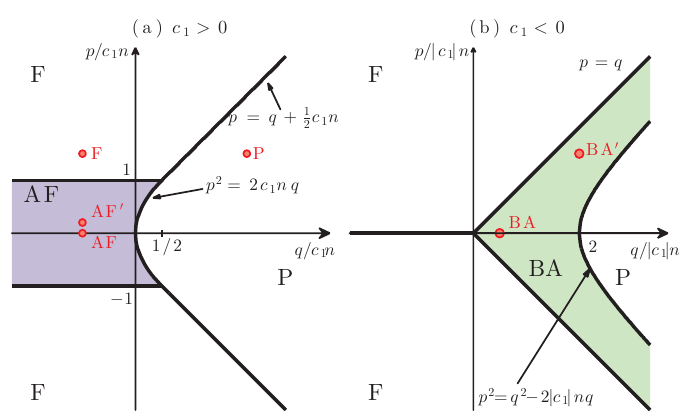} 
\caption{\label{fig:PD}  (Color online) The  phase diagram of a spin-1 condensate with  (a) \emph{antiferromagnetic}  
  and (b) \emph{ferromagnetic} interactions. The vertical
and horizontal axes are the linear and quadratic Zeeman energies in units of the spin-dependent interaction energy $|c_{1}|n$. The phases shown are (F) ferromagnetic, (P) polar, (AF) antiferromagnetic,
 and broken-axisymmetric (BA) (see Refs.~\cite{Stenger1999a,Kawaguchi2012R}).
The rotational symmetry about the direction of the applied field is
spontaneously broken in the AF and BA phases. The particular cases we present results for the dynamic structure factors are indicated on the phase diagram by filled circles labelled: F $(-1,1.5)$,  P $(2.1,1.5)$,  AF $(-1,0)$, AF$^\prime$ $(-1,0.2)$,  BA  $(0.5,0)$ and BA$^\prime$ $(2,1.5)$, where the tuples $(q,p)$ indicate the Zeeman coordinates of the state in units of $|c_1|n$.
}
\end{figure}

\begin{figure*}[!ht] 
\includegraphics[width=6.8in]{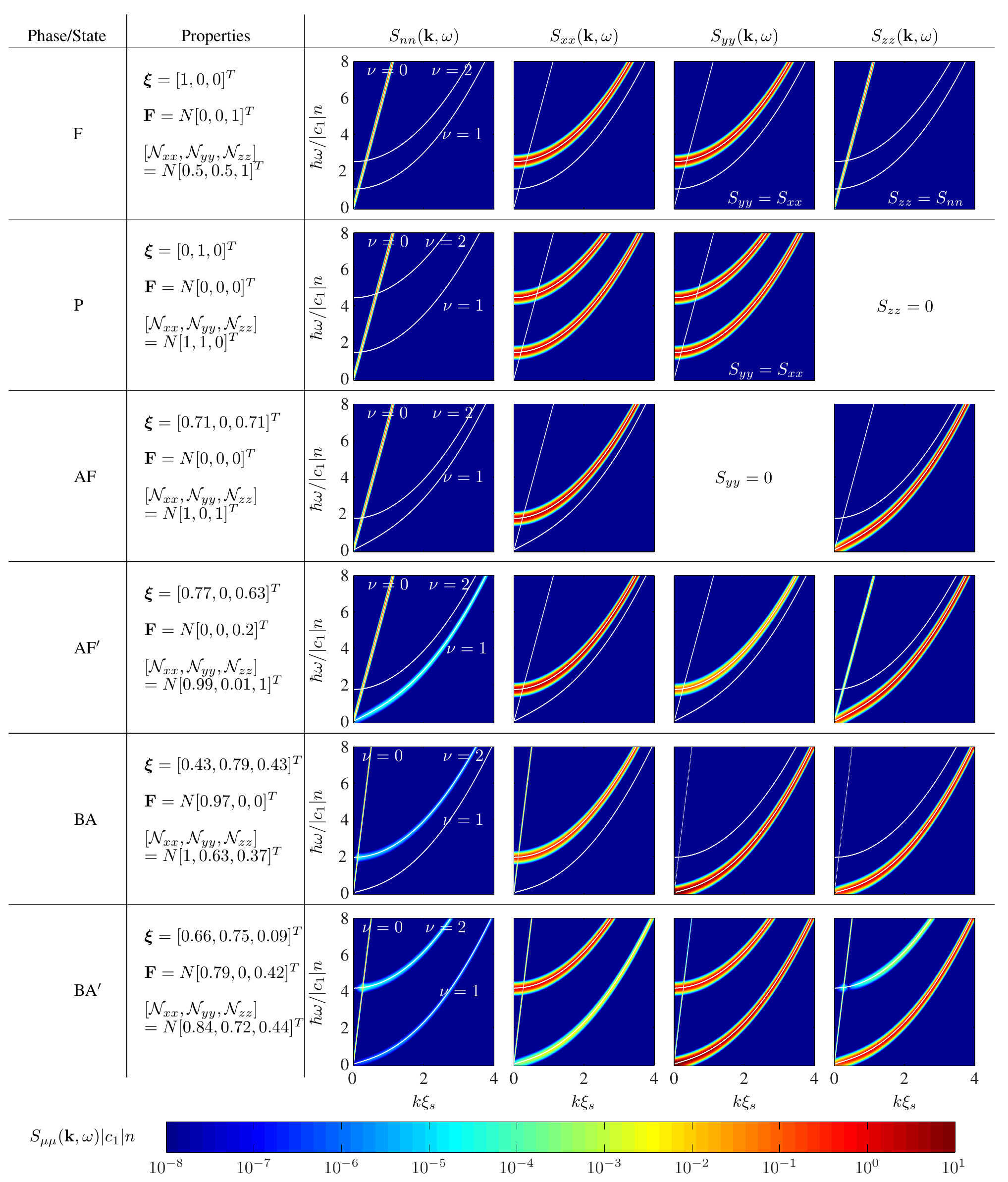}
\caption{\label{fig:dsfall} (Color online) Dynamic structure factor at $T=0$ for the six phases/states indicated in Fig.~\ref{fig:PD}. The $\delta$-functions appearing in the dynamic structure factors are frequency broadened by setting $\delta(\omega) \approx e^{-(\omega/W)^2}/\sqrt{\pi}W$ with $W=c_1n/10\hbar$. The thin white lines indicate the quasiparticle dispersion relations $\omega_{\bk\nu}$ for reference (the $\nu=0$ branch is dotted in the BA and BA$'$ cases for clarity). For the BA results we have chosen interaction parameters relevant to $^{87}$Rb i.e.~$c_0=250\, |c_1|$ (with $c_1<0$), and for all other results we have used $^{23}$Na interaction parameters, i.e.~$c_0=50\, c_1$ (with $c_1>0$). The Zeeman parameters for the various results are indicated in Fig.~\ref{fig:PD}. Wave vectors are scaled by the spin healing length $\xi_s=\hbar/\sqrt{M|c_1|n}$. For the BA phase, $S_{yy}(\bk,\omega)$ is divergent at $k=0$ and the scale has been capped. }
\end{figure*}

In the following subsections we consider the dynamic structure factors for cases within the four distinct phases of the spin-1 system. The phase diagram for this system is shown in Fig.~\ref{fig:PD}, for antiferromagnetic interactions [$c_1>0$ Fig.~\ref{fig:PD}(a)] and ferromagnetic interactions [$c_1<0$ Fig.~\ref{fig:PD}(b)]. The phase diagrams are parameterised by the linear and quadratic Zeeman energies. It should be noted that (following the standard meanfield treatment \cite{Stenger1999a,Kawaguchi2012R}) the linear Zeeman energy is in fact the sum of the normal linear Zeeman energy ($\sim-g\mu_BB$) and   a Lagrange multiplier introduced to constrain the value of the $z$-component of magnetization, which is a constant of motion for the Hamiltonian (\ref{Hspin1}).
The particular cases we consider are labelled $\{$F, P, AF, AF$^\prime$, BA, BA$^\prime\}$  and their corresponding Zeeman parameters are shown as points on the phase diagram (see Fig.~\ref{fig:PD}).
The main results for these phases are collated together in Fig.~\ref{fig:dsfall} (with off-diagonal results in Fig.~\ref{f:offdsf}) to enable easier comparison of the similarities and differences between the phases as revealed by the dynamic structure factors. In these figures we have also shown the dispersion relations for the three quasiparticle branches, labelled $\nu=0,1,2$. We always label phonon branch as $\nu=0$. The phonon branch is gapless in all phases (it is the Nambu-Goldstone mode associated with the broken $U(1)$ symmetric occurring at condensation) and rises steeply since we take the density interaction to be much larger than the magnitude of the spin-dependent interaction (i.e.~$c_0\gg|c_1|$). We have verified that all $f$-sum rules are satisfied for all our results presented. 
The static structure factors for the spin-1 case, including various analytic results, have already been considered in Ref.~\cite{Symes2014a}. $S_{\mu\mu}$ is always real and positive, $S_{\mu'\mu}=S_{\mu\mu'}^*$ and, for the states we consider, 
$S_{nx}$, $S_{nz}$, $S_{xz}$ are real and $S_{ny}$, $S_{xy}$, $S_{yz}$ are purely imaginary.

\subsection{Ferromagnetic phase}
In the ferromagnetic phase the condensate is fully magnetized along $z$, with the condensate residing completely in the $m=+1$ sub-level. The particular state we consider is labelled F in Fig.~\ref{fig:PD}, and the corresponding dynamic structure factors $S_{\mu\mu}$, and other relevant parameters for this state, are shown in Fig.~\ref{fig:dsfall}.

Because the condensate only occupies the $m=+1$ sub-level [see order parameter in Fig.~\ref{fig:dsfall}], the $S_{nn}$ and $S_{zz}$ dynamic structure factors are identical, and only couple to the ($\nu=0$) phonon branch of excitations.  
The $S_{xx}$ and $S_{yy}$ dynamic structure factors are identical. These both couple to the $\nu=2$ magnon branch of transverse magnetic excitations. This excitation branch is a  gapped with the associated Bogoliubov amplitudes residing completely in the $m=0$ sub-level.
For the off-diagonal dynamic structure factors ($S_{\mu\mu'}$), which are not shown: $S_{nz}$ is identical to $S_{nn}$ and $S_{zz}$, $S_{nx}=S_{ny}=S_{xz}=S_{yz}=0$, and, at $T=0$, $S_{xy}=iS_{xx}$. This off-diagonal behavior of $S_{\mu\mu'}$ can be understood with reference to Eq.~\eqref{e:Bogsmumud}: this expression is only non-zero if the same excitation branch contributes to both $\delta\tilde{F}_{\mu,\bk\nu}$ and $\delta\tilde{F}_{\mu',\bk\nu}$ or equivalently $S_{\mu\mu}$ and $S_{\mu'\mu'}$. For the F phase we see from Fig.~\ref{fig:dsfall} that this only occurs for $\delta\tilde{n}_{\bk\nu}$, $\delta\tilde{F}_{z,\bk\nu}$ (both of which only couple to the $\nu=0$ branch) so $S_{nz}$ is non-zero and $\delta\tilde{F}_{x,\bk\nu}$, $\delta\tilde{F}_{y,\bk\nu}$ (both of which couple to the $\nu=2$ branch) so $S_{xy}$ is non-zero.
 We also note that the $\nu=1$ magnon branch is associated with nematic fluctuations, and  does not contribute to any of the dynamic structure factors we consider here. To probe this branch would require a nematic probe that directly couples the $m=+1$ and $m=-1$ sub-levels.

\subsection{Polar phase}
In the polar phase the condensate is in an unmagnetized nematic state, with a nematic director aligned to the $z$-axis \cite{Seo2015a,Zibold2015a}. In this phase the condensate is completely in the $m=0$ sub-level. 
The polar state we consider is labelled P in Fig.~\ref{fig:PD}.   The dynamic structure factors $S_{\mu\mu}$, and other relevant parameters for this state,  are shown in Fig.~\ref{fig:dsfall}. 
In this state  $S_{nn}$  couples only to the $(\nu=0)$ phonon branch. The $\nu=1,2$ branches are both transverse magnetic excitations and contribute to $S_{xx}$ and $S_{yy}$.  Because there is no condensate occupation in the $m=\pm1$ sub-levels the $S_{zz}$ response is zero [by the matrix element in \eqref{e:tildeFdef}] at the level of approximation we consider here.
For the reasons given in the discussion of the ferromagnetic phase, the only off-diagonal structure factor that is non-zero is $S_{xy}$. Since $\delta\tilde{F}_{y,\bk,1} = i\delta\tilde{F}_{x,\bk,1}$ and $\delta\tilde{F}_{y,\bk,2}=-i\delta\tilde{F}_{x,\bk,2}$ we have $S_{xy}=\pm i S_{xx}$ at $T=0$ with $-$ for the lower ($\nu=1$) branch and $+$ for the upper  ($\nu=2$)  branch.

\subsection{Antiferromagnetic phase}

In the antiferromagnetic phase, like the polar phase,  the condensate is in a nematic state (although not necessarily unmagnetized), however with a nematic director that lies in the $xy$-plane. This phase hence breaks the continuous axial spin symmetry of the Hamiltonian and the magnon   branch (labeled $\nu=1$) becomes gapless (i.e.~it is the Nambu-Goldstone mode associated with the broken symmetry). The anti-ferromagnetic phase is partially magnetized along $z$ for $p\ne0$. We consider the $p=0$ and $p\ne0$ cases separately below.

\subsubsection{$p=0$ case}
The $p=0$ anti-ferromagnetic state we consider is labelled AF in Fig.~\ref{fig:PD}.   The dynamic structure factors $S_{\mu\mu}$, and other relevant parameters for this state,  are shown in Fig.~\ref{fig:dsfall}. 
In this state $S_{nn}$  couples only to the $(\nu=0)$ phonon branch, and the  $S_{zz}$ dynamic structure factor only couples to gapless magnon branch ($\nu=1$). We note that the broken spin symmetry is associated with rotations about the $z$-axis, generated by $F_z$.
The $\nu=2$ excitation branch couples to $S_{xx}$ but  does not contribute to $S_{yy}$, which is identically zero within the Bogoliubov treatment. This asymmetry in the $xy$-plane can be understood because in choosing a real condensate order parameter [see details in Fig.~\ref{fig:dsfall}] we selected the director to aligned with the $x$-axis. It is possible to prepare the condensate spin ordered in particular directions in experiment (e.g.~see \cite{Seo2015a}), however in general if the symmetry spontaneously breaks the director will  choose a random direction in the $xy$-plane, and $S_{xx}$ and $S_{yy}$ behaviour will be rotated into each other. We note that the mixed structure factors are all zero for this case.

\begin{figure*}[!ht]
\includegraphics[width=6.8in]{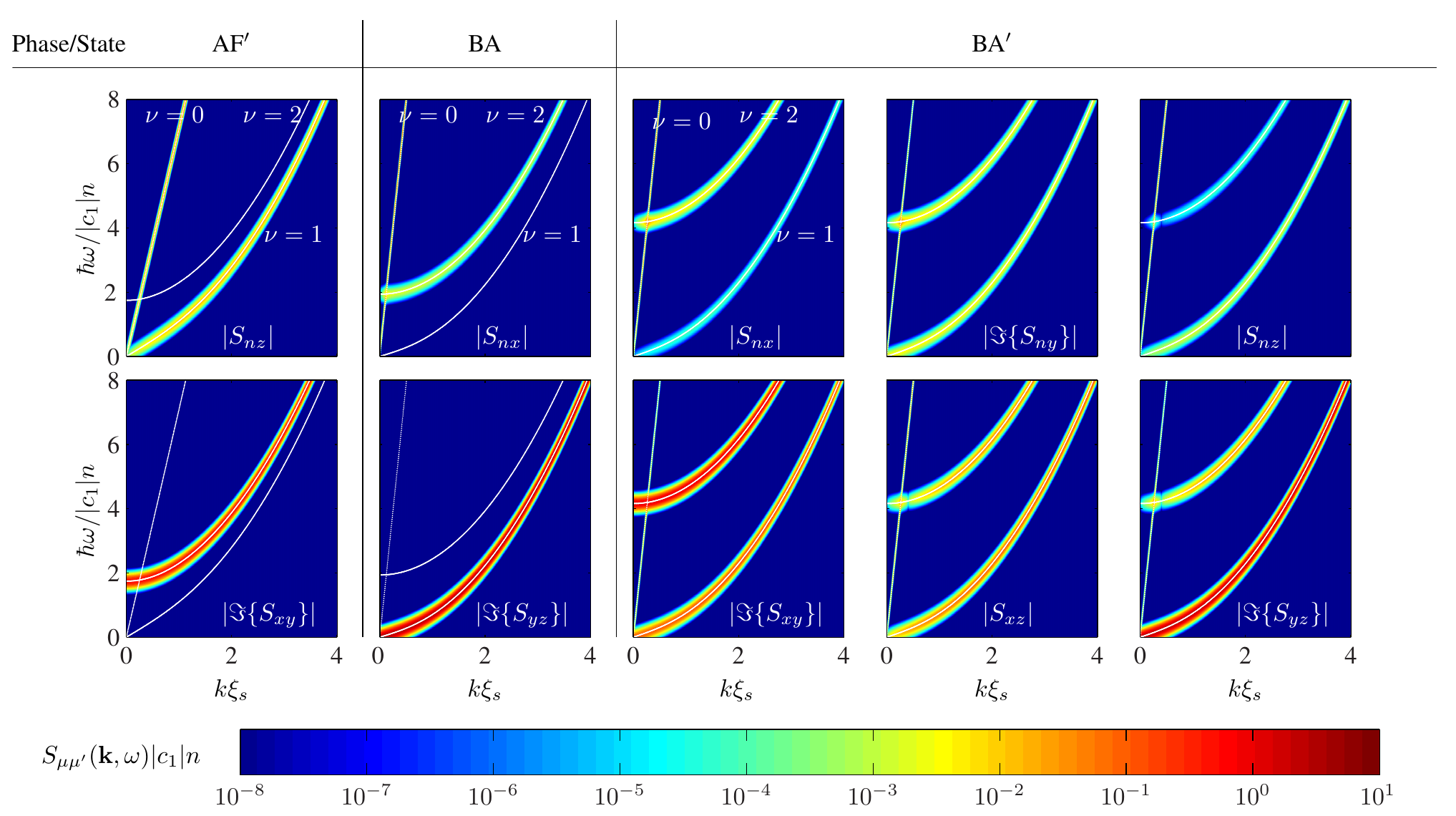}
\caption{\label{f:offdsf} (Color online) Off-diagonal dynamic structure factor at $T=0$. See Fig.~\ref{fig:PD} for details of the phases/states. Other details as in Fig.~\ref{fig:dsfall}.}
\end{figure*}

\subsubsection{$p\ne0$ case}

The $p\ne0$ anti-ferromagnetic state we consider is labelled AF$^\prime$ in Fig.~\ref{fig:PD}, and differs from the AF state by being partially magnetized along $z$.  The dynamic structure factors $S_{\mu\mu}$, and other relevant parameters for this state,  are shown in Fig.~\ref{fig:dsfall}.  A notable difference of this case to the $p=0$ case is that the $\nu=1$ branch now contributes marginally to $S_{nn}$, and similarly the $\nu=0$ branch contributes to $S_{zz}$.  We also observe that  the $\nu=2$ branch couples to both $S_{xx}$ and $S_{yy}$, although more weakly to $S_{yy}$.
Because the $\nu=0$ and $\nu=1$ branches contribute to both $S_{nn}$ and $S_{zz}$ the mixed dynamic structure factor $S_{nz}$ is non-zero  and similarly the $\nu=2$ branch contributes to both $S_{xx}$ and $S_{yy}$ so $S_{xy}$ is non-zero as shown in Fig.~\ref{f:offdsf}.  The other mixed structure factors are zero (i.e.~$S_{nx}=S_{ny}=S_{xz}=S_{yz}=0$).

\subsection{Broken-axisymmetric phase}

The broken-axisymmetric phase only occurs for systems with a ferromagnetic interaction when the quadratic Zeeman energy is positive, but sufficiently small (see Fig.~\ref{fig:PD}). In this phase the condensate is partially magnetized, but the magnetization tilts relative to the $z$-axis. Thus this phase also breaks a continuous symmetry of the Hamiltonian, with an associated Nambu-Goldstone magnon mode emerging (the excitation branch labelled $\nu=1$).
Experiments have extensively investigated the spontaneous symmetry breaking dynamics of this phase  (e.g.~see \cite{Sadler2006a}).
For $p=0$ the magnetization lies in the $xy$-plane while for $p\ne0$ it tilts out of plane. We consider these two cases separately in what follows.

\subsubsection{$p=0$ case}

The $p=0$ broken-axisymmetric  state we consider is labelled BA in Fig.~\ref{fig:PD}, and the dynamic structure factors $S_{\mu\mu}$ for this phase are shown in Fig.~\ref{fig:dsfall}.
In this state $S_{nn}$ couples to the phonon ($\nu=0$) and the gapped magnon  ($\nu=2$) branches. The $\nu=2$ contribution appears to suddenly turn on at small $k$ where the $\nu=0$ dispersion intersects with the $\nu=2$ dispersion (indeed, an avoided crossing occurs between these branches). The $S_{zz}$ response couples only to the gapless magnon ($\nu=1$) branch. We note that the broken spin symmetry is associated with rotations about the $z$-axis, generated by $F_z$.
The $S_{xx}$ and $S_{yy}$ dynamic structure factors are asymmetric in the sense that they couple to the $\nu=2$ and $\nu=1$ branches, respectively ($S_{xx}$ also couples to $\nu=0$). The particular asymmetry we observe here arises because we have chosen a real condensate order parameter with the consequence that the magnetization lies along the positive $x$-axis.
For the ($p=0$) BA state only the off-diagonal dynamic structure factors $S_{nx}$ and $S_{yz}$ are non-zero and are shown in Fig.~\ref{f:offdsf}, while $S_{ny}=S_{nz}=S_{xz}=S_{xy}=0$.

\subsubsection{$p\ne0$ case}
The $p\ne0$ broken-axisymmetric  state we consider is labelled BA$'$ in Fig.~\ref{fig:PD} and the dynamic structure factors $S_{\mu\mu}$ for this phase are shown in Fig.~\ref{fig:dsfall}.
This state is a partially magnetized state, and for $p\ne0$ this magnetization has a non-zero $z$-component.   Also, for $p\ne0$ the $\nu=0$ and $\nu=1$ branches exchange some of their character  (this behavior of the $\nu=1$ branch is referred to as the phonon-magnon coupled mode in Ref.~\cite{Murata2007a}), i.e.~(as compared to the $p=0$ case) the $\nu=1$ mode couples to $S_{nn}$ and the $\nu=0$ branch couples to $S_{zz}$. Indeed, for this case we find that all three branches couple to each of the dynamic structure factors $S_{\mu\mu}$. This means that all of the mixed structure factors $\{S_{nx},S_{ny},S_{nz},S_{xy},S_{xz},S_{yz}\}$ are non-zero, and are shown in Fig~\ref{f:offdsf}.
 
\section{Conclusions and outlook}
Here we have developed the theory for Bragg spectroscopy of spinor Bose-Einstein condensates, notably allowing for the possibility of spin-coupled probing as has been realised in recent experiments with Fermi gases \cite{Hoinka2012a}. We consider both the total momentum and energy imparted is used as observables, and we showed that within the linear response regime, these relate to various density and spin-density dynamic structure factors. Using exact commutation relations of the many-body spinor Hamiltonian we derived the $f$-sum rule, which provides a rigorous constrain on the first frequency moment of the these dynamic structure factors. We find that the $f$-sum rule in general depends on the Zeeman fields, the $z$-magnetization and the diagonal elements of the nematic tensor of the system.

 We have developed expressions for the various dynamic structure factors assuming a meanfield condensate and a Bogoliubov description of the quasi-particle excitations. This should provide a good description of the system at temperatures well below the condensation temperature.  We numerically evaluated the dynamic structure factors for a spin-1 condensate to illustrate their sensitivity to the different phonon and magnon excitation branches. The character of these excitation branches, and how they contribute to the dynamic structure factors, changes significantly between the phases. These results demonstrate that Bragg spectroscopy gives access to properties of spinor condensates that have not yet been directly probed in experiments.

For scalar condensates the static structure factor has also been measured  using high resolution \textit{in situ} density measurements (e.g.~see \cite{Hung2011b}). By employing magnetization sensitive imaging (e.g.~see \cite{Higbie2005a}) it
may be possible to extend such schemes to the static spin structure factors \cite{Symes2014a}. However, unlike the Bragg spectroscopy approach, such measurements are sensitive to thermal effects. Because the typical size of the spin-dependent energy is small, thermal effects will be very significant (i.e.~experiments will typically be in the temperature regime $k_BT\gg |c_1|n$). 

In this work we have restricted our attention to probing that couples to the density and components of spin. As future work it would be interesting to devise a scheme to directly couple to nematic densities, known to be important in the complete description of spinor condensates (e.g.~see \cite{Yukawa2012a,Zibold2015a}).

\appendix

\section*{Acknowledgments}
We gratefully acknowledge support by the Marsden Fund of the Royal Society of New Zealand (contract number UOO1220). We would like to acknowledge valuable discussions with L.~Turner and R.~Anderson.

\appendix
\section{Linear response\label{s:linresp}}

The dynamic susceptibility $\chi_{F,B^\dagger}(\bk,\omega)$ describes the evolution of an operator $\Fh_\bk$ under the effect of the perturbation given in Eq.~(\ref{eq:HBragg}) according to
\begin{align}
    \delta\av{\Fh_\bk} &\equiv -\frac{V}2   e^{-i\omega t}e^{\eta t} \chi_{F,B^\dagger}(\bk,\omega) - \frac{V}2  e^{i\omega t}e^{\eta t} \chi_{F,B}(\bk,-\omega).
\end{align}
The susceptibility is given by the standard result \cite{BECbook}
\begin{align}
    \chi_{F,B^\dagger}(\bk,\omega)  
    &= -\sum_{mn} \frac{e^{-\beta E_m}}{\hbar\z}\Biggl[\frac{\av{m|\Fh_\bk|n}\av{n|\Bh^\dagger_\bk|m}}{ \omega - \omega_{nm}+i\eta }\notag\\
    &- \frac{\av{m|\Bh^\dagger_\bk|n}\av{n|\Fh_\bk|m}}{ \omega + \omega_{nm}+i\eta }  \Biggr].
\end{align}
Using the definition \eqref{Eq:Smu} we have the relationship between the dynamic susceptibility and the dynamic structure factors
\begin{align}
  \chi_{F,B^\dagger}(\bk,\omega) 
  &=  \intinf d\omega' \frac{e^{i(\omega-\omega')t }-1}{ \omega - \omega' }\notag\\
  &\times\left[ S_{F,B^\dag}(\bk,\omega')   - S_{B^\dag,F}(\bk,-\omega') \right].
  \end{align}
Ignoring terms rotating at $2\omega$
\begin{align}
    \delta\av{\Bh_\bk} &=  -\frac{V}2 \intinf d\omega' \frac{e^{-i\omega' t }-e^{-i\omega t}}{ \omega - \omega' }\notag\\
      &\times\left[ S_{B,B^\dag}(\bk,\omega')   - S_{B^\dag,B}(\bk,-\omega') \right],\\
\tdiff{P_{z}(t)}{t}  &=  -M\omega_z^2 Z - \frac{ikV}{2} \left( \av{\delta\Bh_\bk^\dag} e^{-i\omega t} - \cc\right),\\
    \tdiff{E(t)}{t}  &= -\frac{V}2 \left(e^{-i\omega t} \tdiff{\av{\Bh^\dag_\bk}}{t} + e^{i\omega t} \tdiff{\av{\Bh_\bk}}{t}\right),
\end{align}
giving Eqs.~\eqref{EqPresponse} and \eqref{EqEresponse}.
\section{Commutators for $f$-sum rule \label{s:sumrules}}
In this Appendix we present the commutation results for the spinor Hamiltonian (\ref{Hspinor}) used to derive the $f$-sum rules.
We introduce the following notation:
\begin{align}
    \rhoh_{mm',\bk} &=\int d\bx\,e^{-i\bk\cdot\bx}\psihd_{m}(\bx)\psih_{m'}(\bx),\\
    \Nh_{mm'} &= \int d\bx\, \psihd_{m}(\bx)\psih_{m'}(\bx), \hspace{3mm} N_{mm'} = \av{\Nh_{mm'}},\\
    \Ah(\bx) &= \sum_{mm'}A_{mm'}\psih^\dagger_m(\bx)\psih_{m'}(\bx),\label{e:opform}\\
    \Ah &=\int d\bx\,\Ah(\bx) = \sum_{mm'} A_{mm'} \Nh_{mm'},\\
    \Ah_{\bk} &= \int d\bx\,e^{-i\bk\cdot\bx} \Ah(\bx) = \sum_{mm'} A_{mm'} \rhoh_{mm',\bk}.
 \end{align}
\subsection{Spin-independent single particle Hamiltonian}
Since the trap is independent of spin, only the kinetic term contributes, giving
\begin{align}
    \av{[\rhoh_{nn',\bk}, [\Hh_0, \rhoh_{mm',-\bk}]]} 
 &=  \frac{\hbar^2k^2 }{2M} \left(N_{nm'}\delta_{n'm} + N_{mn'}\delta_{nm'}\right),
\end{align}
where we assumed there is no spin current. 
Then
\begin{align}
    \av{[\Ah_{\bk}, [\Hh_0, \Bh_{-\bk}]]} \!&= \hspace{-3.5mm} \sum_{nn'mm'}  \hspace{-3mm} A_{nn'} B_{mm'}\av{[\rhoh_{nn',\bk}, [\Hh_0, \rhoh_{mm',-\bk}]]},\\
  &= \frac{\hbar^2 k^2}{2M}\av{\widehat{AB} + \widehat{BA}},
\end{align}
which gives the kinetic term in \eqref{e:sumrules}. 

    \subsection{Zeeman term}
    For any operator $\Zh(\bx)$ of the form \eqref{e:opform}
\begin{align}
 \av{[\rhoh_{mm',\bk}, [\Zh, &\rhoh_{nn',-\bk}]]} =\sum_{l} \Big[ Z_{ln} (N_{mn'}\delta_{m'l}  - N_{lm'}\delta_{mn'}) \notag\\
 &-  Z_{n'l}(N_{ml}\delta_{m'n}  - N_{nm'}\delta_{ml})\Big],
 \end{align}
so $\av{[\Ah_{\bk}, [\Zh,\Bh_{-\bk}]]} = \av{[\Ah,[\Zh,\Bh]]}$. Setting $\Zh\to\Hh_Z=-p\Fh_z+q\hat{\N}_{zz}$ gives \eqref{e:m1zeeman}.
    \subsection{Interaction terms}
Setting ${}^\F C_{m'n'}^{mn} = \sum_{\M} \av{mn|\F\M}\av{\F\M|m'n'}$ and $\Vh^{mn}_{m'n'} \equiv \int d\bx\,\psihd_m(\bx)\psihd_{n}(\bx)\psih_{m'}(\bx)\psih_{n'}(\bx)$, the interaction terms (\ref{spinorints}) can be written
\begin{align}
    \Hint^{(\F)} &= \frac{g_\F}{2} \sum_{mnm'n'} \hspace{-2mm}{}^\F C_{m'n'}^{mn}\Vh^{mn}_{m'n'}.
\end{align}
The commutation expression for this is
\begin{align}
    \av{[\Ah_{\bk},[\Hint^{(\F)},\Bh_{-\bk}]]} 
    &= \frac{g_\F}{2}\sum_{mnm'n'} {}^\F C_{m'n'}^{mn}    \av{[\Ah,[ \Vh_{m' n'}^{mn} ,\Bh]] }.
\end{align}

The spinor interaction is spherically symmetric, so we may arbitrarily choose the $z$ direction \cite{Kudo2010a} to find
\begin{align}
    &[\Hint^{(\F)},\Fh_z] = \frac{g_\F}{2}\sum_{mnm'n'} {}^\F C_{m'n'}^{mn} \sum_{l} l [ \Vh_{m' n'}^{mn} ,\Nh_{ll}],\\
&=  \frac{g_\F}{2}\hspace{-2mm}\sum_{\M mnm'n'}\hspace{-4mm}  \av{mn|\F\M}\av{\F\M|m'n'} \Vh_{m'n'}^{nm}( n'\! +\! m'\! -\! n\!-\!m),
\end{align}
which is zero using $m+n=m'+n'=\M$. We note that Ref.~\cite{Kunimi2015a} shows an alternative proof that the commutator of the interaction Hamiltonian is zero.
%

%

\end{document}